# $\Theta^+$ in a conventional correlated perturbative chiral quark model (CP$\chi$QM )

A. R. Haghpayma†

*Department of Physics, Ferdowsi University of Mashhad*

*Mashhad, Iran*

## Abstract

Assuming a conventional correlated perturbative chiral quark model (CP$\chi$QM ) , we suggest that the $\Theta^+$ baryon is a bound state of two vector diquarks and a single antiquark, the spatially wave function of these diquarks has a P- wave and a S- wave in angular momentum in the first and se - cond version of our model respectively, as the result of these considerations we construct the orbi - tal - colour - flavour - spin symmetry of $q^4\bar{q}$ contribution of quarks and by imposing considerations such as HF interactions and suitable diquark currents we leads to a conventional pentaquark inte - rpolating field which may be used for QSR analysis. In order to get an idea of the general features of the spectrum we have calculated the mass spectrum of exotic pentaquark states with the Gürsey Radicati mass formula.

Pentaquark baryons may be pure exotic or Crypto- exotic the pure exotic states can easily be identified by their unique quantum numbers, but the crypto - exotic states are hard to be identified as their quantum numbers can also be generated by three - quark states. therefore, it is crucial to have careful analyses for their decay channels.

Quark models have provided a cornerstone for hadron physics and one can studying the structure of pentaquark baryons in a naive constituent quark model [1] .

In Ref [2] karliner and lipkin suggested a triquark - diquark model where , for example, $\Theta^+$ is a system of ( ud ) - ( ud$\bar{s}$).

In Ref [3] jaff and wilczek presented a diquark - diquark- antiquark model so $\Theta^+$ is ( ud ) - ( ud ) - $\bar{s}$ , in this model they also considered the mixing of the pentaquark antidecuplet with the pentaquark octet, which makes it different from the SU ( 3 ) soliton models where the octet describes the normal three- quark baryon octet.

More predictions based on quark models can be found in Ref [1].

In order to investigate mass, width, reaction channels and other properties of pentaquarks a set of its wave functions in quark model is requierd [4] .

By introducing the quark and antiquark operators and by taking direct product of two diquarks and one antiquark the $\Theta^+$ state can be represented by SU ( 3 ) tensors.

The $SU_f$ ( 3 ) symmetric lagrangian for pentaquark baryons and their interactions with other multiplets can be constructed by imposing dynamical interactions between quarks , the symmetry breaking can be included.

We denote a quark with $q_i$ and antiquark with $q^i$ in which $i$ = 1,2,3 denote u , d , s and impose normalization relations as : $(q_i,q_j)=\delta_{ij}$, $(q^i,q^j)=\delta^{ij}$, $(q_i,q^j)=0$ and in $(p\ ,\ q)$ notation we consider a tensor $T^{b_1,\dots,b_q}_{a_1,\dots,a_p}$ which is completely symmetric in upper and lower indices and traceless on every pair of indices.

We introduce $S_{jk}=\frac{1}{\sqrt{2}}(q_jq_k+q_kq_j)$ and $A_{jk}=\frac{1}{\sqrt{2}}(q_jq_k-q_kq_j)$ then for a quark and an antiquark we would have

quark : 3 : $T_i$ $i=1,2,3$

antiquark : $\bar{3}$ : $T^i=\epsilon^{ijk}A_{jk}$, $ijk=1,2,3$

we have:

$$(T^i,T^j)=4\delta^{ij}.\ (T_i,T_j)=4\delta_{ij}.\ (S_{jk},S_{lm})=\delta_{jl}\delta_{km}+\delta_{jm}\delta_{kl},\quad (S_{jk},T^i)=0 \tag{1}$$

for $q^4$ we have:

$$(\mathbf{3}\otimes\mathbf{3}\otimes\mathbf{3}\otimes\mathbf{3})=(\mathbf{6}\oplus\bar{\mathbf{3}})\otimes(\mathbf{6}\oplus\bar{\mathbf{3}})=(\mathbf{6}\otimes\mathbf{6})\oplus(\mathbf{6}\otimes\bar{\mathbf{3}})\oplus(\bar{\mathbf{3}}\otimes\mathbf{6})\oplus(\bar{\mathbf{3}}\otimes\bar{\mathbf{3}})$$
$$=(\mathbf{15}_1\oplus\mathbf{15}_2\oplus\bar{\mathbf{6}})\oplus(\mathbf{15}_2\oplus\mathbf{3})\oplus(\mathbf{15}_1\oplus\mathbf{3})\oplus(\bar{\mathbf{6}}\oplus\mathbf{3}) \tag{2}$$

and for $q^4\bar{q}$ we have:

$$[\mathbf{3}\otimes\mathbf{3}\otimes\mathbf{3}\otimes\mathbf{3}]\otimes\bar{\mathbf{3}}=([\mathbf{15}_1\oplus\mathbf{15}_2\oplus\bar{\mathbf{6}}]\oplus[\mathbf{15}_2\oplus\mathbf{3}]\oplus[\mathbf{15}_1\oplus\mathbf{3}]\oplus[\bar{\mathbf{6}}\oplus\mathbf{3}])\otimes\bar{\mathbf{3}}$$
$$=2(\mathbf{15}_1\otimes\bar{\mathbf{3}})\oplus2(\mathbf{15}_2\otimes\bar{\mathbf{3}})\oplus2(\bar{\mathbf{6}}\otimes\bar{\mathbf{3}})\oplus2(\mathbf{3}\otimes\bar{\mathbf{3}})$$
$$=2(\mathbf{35}\oplus\mathbf{10})\oplus2(\mathbf{27}\oplus\mathbf{10}\oplus\mathbf{8})\oplus2(\overline{\mathbf{10}}\oplus\mathbf{8})\oplus2(\mathbf{8}\oplus\mathbf{1})$$

The tensor notation of all of these multiplets can be constructed as $T_i$ and $T^i$.

For example the tensor notations for $\overline{10}$ and 8 are [5] :

$$\overline{10}:\quad T^{ijk}=\frac{c_1}{\sqrt{3}}\left(S^{ij}\bar{q}^k+S^{jk}\bar{q}^i+S^{ki}\bar{q}^j\right)+\frac{c_2}{\sqrt{3}}\left(T^{ij}\bar{q}^k+T^{jk}\bar{q}^i+T^{ki}\bar{q}^j\right), \tag{3}$$

$$8:\quad P^i_j=\frac{c_1}{\sqrt{2}}\left(T_j\bar{q}^i-\frac{1}{3}\delta^i_jT_m\bar{q}^m\right)+\frac{c_2}{\sqrt{2}}\left(Q_j\bar{q}^i-\frac{1}{3}\delta^i_jQ_m\bar{q}^m\right)+\frac{c_3}{\sqrt{2}}\left(\widetilde{Q}_j\bar{q}^i-\frac{1}{3}\delta^i_j\widetilde{Q}_m\bar{q}^m\right)$$
$$+\frac{c_4}{\sqrt{3}}\epsilon_{jab}S^{ia}\bar{q}^b+\frac{c_5}{\sqrt{3}}\epsilon_{jab}T^{ia}\bar{q}^b+\frac{c_6}{\sqrt{15}}T^i_{jk}\bar{q}^k+\frac{c_7}{\sqrt{15}}\widetilde{T}^i_{jk}\bar{q}^k+\frac{c_8}{\sqrt{15}}S^i_{jk}\bar{q}^k. \tag{4}$$

$$\widetilde{Q}_m=\frac{1}{\sqrt{8}}T^lS_{ml},\qquad \widetilde{T}^i_{jk}=T^iS_{jk}-\frac{1}{\sqrt{2}}\left(\delta^i_j\delta^m_k+\delta^i_k\delta^m_j\right)\widetilde{Q}_m.$$
$$T^{ij}=\frac{1}{\sqrt{6}}\epsilon^{iab}\epsilon^{jcd}S_{ac}S_{bd},\qquad \widetilde{Q}_m=\frac{1}{\sqrt{8}}T^lS_{ml},$$
$$S^i_{jk}=\frac{}{\sqrt{2}}\epsilon^{ilm}\left(S_{jl}S_{km}+S_{kl}S_{jm}\right),\qquad T^i_{jk}=S_{jk}T^i-\frac{1}{\sqrt{2}}\left(\delta^i_j\delta^m_k+\delta^i_k\delta^m_j\right)Q_m,$$

The nomenclature for pentaquark states based on hypercharge is as follow: $Y=2$ $\Theta$, $Y=1$ $N,\Delta$, $Y=0$ $\Sigma,\Lambda$, $Y=-1$ $\Xi$ , $Y=-2$ $\Omega$, $Y=-3$ $X$,

† e-mail: haghpeima@wali.um.ac.ir

which in the notation $T^{b_1,\ldots,b_q}_{a_1,\ldots,a_p}$ we have[6] :

$$Y = p_1 - q_1 + p_2 - q_2 + \frac{3}{2}(p-q),$$

$$I_3 = \frac{1}{2}(p_1 - q_1) - \frac{1}{2}(p_2 - q_2). \qquad p_1 + p_2 + p_3 = p \text{ and } q_1 + q_2 + q_3 = q$$

$$Q = I_3 + Y/2$$

We can constract the $SU_f(3)$ symmetry lagrangians for example:

$$\mathcal{L}_{\overline{\mathbf{10}}\text{-}\mathbf{8}_3} = g_{\overline{\mathbf{10}}\text{-}\mathbf{8}_3}\,\epsilon^{ilm}\overline{T}_{ijk}B^j_l M^k_m + (\text{H.c.}). \tag{5}$$

And mass terms, for example:

$$H_{\mathbf{8}} = a\overline{P}^i_j P^j_i + b\overline{P}^i_j \mathcal{Y}^l_i P^j_l + c\overline{P}^i_j \mathcal{Y}^j_l P^l_i. \tag{6}$$

$$H_{\overline{\mathbf{10}}} = a\overline{T}_{ijk}T^{ijk} + b\overline{T}_{ijk}\mathcal{Y}^k_l T^{ijl}, \tag{7}$$

for pentaquark octet and antidecuplet respectively.

The pentaquark wave function contains contributions connected to the spatial degrees of freedom and the internal degrees of freedom of color, flavour and spin, the pentaquark wave function should be a color singlet as all physical states, and should be antisymmetric under permutation of the four quarks ( pauli principle ) .

In order to classify quark and antiquark in $SU_f(3)$ , $SU_s(2)$ we introduce the notations:

| | $SU_{sf}(6)$ | $\supset$ | $SU_f(3)$ | $\otimes$ | $SU_s(2)$ |
|---|---|---|---|---|---|
| quark | [1] | $\supset$ | [1] | $\otimes$ | [1] |
| | □ | $\supset$ | □ | $\otimes$ | □ |
| antiquark | [11111] | $\supset$ | [11] | $\otimes$ | [1] |
| | □<br>□<br>□<br>□<br>□ | $\supset$ | □<br>□ | $\otimes$ | □ |

Thus by taking the outer product of quark and antiquark representations we have:

For $q^4$ $SU_{fs}(6)$:

$$[\,[1]_6 \otimes [1]_6 \otimes [1]_6\,] \otimes [1]_6 = (\,[3] \oplus 2\,[21]_{70} \oplus [111]_{20}\,) \otimes [1]_6 \tag{8}$$

$$= (\,[3]_{56} \otimes [1]_6\,) \oplus 2\,(\,[21]_{70} \otimes [1]_6\,) \oplus (\,[111]_{20} \otimes [1]_6\,)$$

$$= (\,[3]_{56} \otimes [1]_6\,) \oplus 2\,(\,[21]_{70} \otimes [1]_6\,) \oplus (\,[111]_{20} \otimes [1]_6\,)$$

$$= (\,[4]_{126} \oplus [31]_{210}\,) \oplus 2\,(\,[31]_{210} \oplus [22]_{105} \oplus [211]_{105}\,)$$

$$\oplus (\,[211]_{105} \oplus [1111]_{15}\,).$$

$$= [4]_{126} \oplus 3\,[31]_{210} \oplus 2\,[22]_{105} \oplus 3\,[211]_{105} \oplus [1111]_{15}$$

One can decompose these multiplets:

Spin-flavour decomposition of $q^4$ states

| $D_3$ | $\mathcal{T}_d$ | $SU_{sf}(6)$ | $\supset$ | $SU_f(3)$ | $\otimes$ | $SU_s(2)$ |
|---|---|---|---|---|---|---|
| $A_1$ | $A_1$ | $[4]_{126}$ | | $[4]_{15}$ | $\otimes$ | $[4]_5$ |
| | | | | $[31]_{15}$ | $\otimes$ | $[31]_3$ |
| | | | | $[22]_6$ | $\otimes$ | $[22]_1$ |
| $A_1+E$ | $F_2$ | $[31]_{210}$ | | $[4]_{15}$ | $\otimes$ | $[31]_3$ |
| | | | | $[31]_{15}$ | $\otimes$ | $[4]_5$ |
| | | | | $[31]_{15}$ | $\otimes$ | $[31]_3$ |
| | | | | $[31]_{15}$ | $\otimes$ | $[22]_1$ |
| | | | | $[22]_6$ | $\otimes$ | $[31]_3$ |
| | | | | $[211]_3$ | $\otimes$ | $[22]_1$ |
| | | | | $[211]_3$ | $\otimes$ | $[31]_3$ |
| $E$ | $E$ | $[22]_{105}$ | | $[4]_{15}$ | $\otimes$ | $[22]_1$ |
| | | | | $[31]_{15}$ | $\otimes$ | $[31]_3$ |
| | | | | $[22]_6$ | $\otimes$ | $[4]_5$ |
| | | | | $[22]_6$ | $\otimes$ | $[22]_1$ |
| | | | | $[211]_3$ | $\otimes$ | $[31]_3$ |
| $E+A_2$ | $F_1$ | $[211]_{105}$ | | $[31]_{15}$ | $\otimes$ | $[31]_3$ |
| | | | | $[31]_{15}$ | $\otimes$ | $[22]_1$ |
| | | | | $[22]_6$ | $\otimes$ | $[31]_3$ |
| | | | | $[211]_3$ | $\otimes$ | $[4]_5$ |
| | | | | $[211]_3$ | $\otimes$ | $[31]_3$ |
| | | | | $[211]_3$ | $\otimes$ | $[22]_1$ |
| $A_2$ | $A_2$ | $[1111]_{15}$ | | $[22]_6$ | $\otimes$ | $[22]_1$ |
| | | | | $[211]_3$ | $\otimes$ | $[31]_3$ |

For $q^4\bar{q}$ $SU_{fs}(6)$ :

$$[1]_6 \otimes [1]_6 \otimes [1]_6 \otimes [1]_6 \otimes [11111]_6 = [51111]_{700} \oplus 4\,[411111]_{56} \oplus 3\,[42111]_{1134}$$
$$\oplus 8\,[321111]_{70} \oplus 2\,[33111]_{560} \oplus 3\,[32211]_{540}$$
$$\oplus 4\,[222111]_{20} \oplus [22221]_{70}\,. \tag{9}$$

For $q^4\bar{q}$ $SU_f(3)$ :

$$[1]_3 \otimes [1]_3 \otimes [1]_3 \otimes [1]_3 \otimes [11]_3 = [51]_{35} \oplus 3\,[42]_{27} \oplus 2\,[33]_{\overline{10}}$$
$$\oplus 4\,[411]_{10} \oplus 8\,[321]_8 \oplus 3\,[222]_1\,. \tag{10}$$

For $q^4\bar{q}$ $SU_s(2)$ :

$$[1]_2 \otimes [1]_2 \otimes [1]_2 \otimes [1]_2 \otimes [1]_2 = [5]_6 \oplus 4\,[41]_4 \oplus 5\,[32]_2\,, \tag{11}$$

This results are in agreement with the reduction of the color-spin $SU_{cs}(6)$ algebra of Ref [7].

The spin - flavour part has to be combined with the color part and orbital part in such a way that the total pentaquark wave function is a $[222]_1$ color - singlet state, and that the four quarks obey the pauli principle, i.e. are antisymmetric under any permutation of the four quarks, because of the color wave function of an antiquark that is a $[11]_3$ anti -triplet, the color wave function of the four- quark configuration is a $[211]_3$ triplet , the total $q^4$ wave function is antisymmetric, hence the orbital - spin - flavour part is a $[31]$ state which is obtained from the color part by interchanging rows and columns, now if 4 - quarks are in a p - wave state, there are several allowed $SU_{fs}(6)$ representations which are:

[4],[31],[22],[211]

Thus the total orbital - color - flavor - spin wave function of $q^4$ is $[1111]_{ocfs}$

For the explicit pentaquark wave functions see Ref [8].

Now we can impose an specific dynamical model to this general constituent quark model.

Assuming a conventional correlated perturbative chiral quark model (CP$\chi$QM ) , we suggest that the $\Theta^+$ baryon is a bound state of two diquarks and a single antiquark , the spatially wave function of these diquarks has a p - wave and a s - wave in angular momentum in the first and second version of our model respectively.

The $[2]^f$ flavour symmetry of each diquark leads to $[22]^f_{\bar{6}}$ flavour symmetry for $q^4$, the $[2]^s$ spin symmetry of each diquark leads to $[22]^s$ and $[31]^s$ spin symmetry for $q^4$ in the first version and second version of our model respectively.

The color symmetry of each diquark is $[11]^c$ and for the first version of our model we assume $[2]^c$ for one of the diquark pairs this leads to $[211]^c$ color symmetry for $q^4$.

The orbital symmetry of each diquark is $[2]^o$ and for the first version we assume $[11]^o$ for one of the diquark pairs, this leads to $[31]^o$ and $[4]^o$ orbital symmetry for $q^4$ in the first and second version of our model respectively.

As the result of these considerations we would have for $q^4$ contribution of quarks ,( $[1111]^{oc}$ and $[4]^{fs}$ ) and ( $[211]^{oc}$, $[31]^{fs}$) for the first and second versions ; and $[1111]^{ocfs}$ for $q^4$ is the same for two versions and lead to a totally antisymmetric wave function for $q^4$ due to pauli principle.

The flavour symmetry contribution for $q^4$ configuration comes from the decomposition formula (2) $q^4\bar{q}$ ,in which we have two $\bar{6}_f$ the second one is used in J W model and the first one is used in our model due to vector diquark contribution of it.

Thus the tensor notations for $\overline{10}_f$ in our model comes from Eq (3) with ($C_1=0$ and $C_2 \neq 0$ ) also the tensor notations for $8_f$ in our model comes from Eq (4) with $C_i = 0$ $i \neq 5$ , $C_5 \neq 0$ ).

Thus inserting $\overline{10}$ tensor notation into $\mathcal{L}_{\overline{\mathbf{10}}\text{-}\mathbf{8}_3}$ symmetry lagrangian leads to $SU_f(3)$ symmetry interactions which experimental evidence of them would be explored.

Morever inserting $8_f$ and $\overline{10}_f$ tensor notations into conventional quark model mass formulas (6) and (7) respectively leads to $\overline{10}_f$ and $8_f$ mass terms and spacing rules for the introduced particles in these multiplets in terms of experimental parameters ( a , b , and c ) of model.

In our model the diquarks are in $6_f$ and $\bar{3}_c$ symmetry configurations and the hyperfine interactions (color - spin and flavour - spin ) leads to a diquark mass that is greater than the masses which predicted in the models that the diquarks are in $\bar{3}_f$ and $3_c$ symmetry configurations.

Thus the difference between the mass of a diquark and an antiquark in our model is larger than the models which based on scalar diquark instead of vector ones, and thus leads to the current experimental perspective in which no one has found the supersymmetry partners of pentaquark states.

In our model we have used diquark ideas in the chiral limit diquark correlations in the relativistic region . the first version of model yelds the positive parity for pentaquarks and this is in agree withl lattice calculations in which there are attempts to find that states which have the maximom overlape with the positive parity.

In addition to usual HF interactions between quarks in a diquark, these interactions exist between the antiquark $\bar{s}$ and each of the diquarks due to vector configuration of them in our model , these interactions are absent in the scalar diquark models in which there is only electromagnetic interaction between antiquark and each diquark in a pentaquark.

By imposing considerations such as these HF interactions and suitable diquark currents we leads to a conventional pentaquark interpolating field which may be used for QSR analysis.

For example for a diquark with $[\,2\,]^{s}$ and $[\,2\,]^{f}$ spin - flavour configurations and $\ell = 0$ angular momentum one can use:

$$\frac{1}{\sqrt{2}}\left[u^T(x)Cd(x)+d^T(x)Cu(x)\right];\ u^T(x)Cu(x);\ d^T(x)Cd(x), \tag{12}$$

and if the diquark angular momentum is $\ell = 1$ one can use:

$$\frac{1}{\sqrt{2}}\left[u^T(x)C\gamma_\mu d(x)+d^T(x)C\gamma_\mu u(x)\right];u^T(x)C\gamma_\mu u(x);d^T(x)C\gamma_\mu d(x). \tag{13}$$

as diquark currents.

Thus for the first and second version of our model we can consider the following interpolating fields respectively.

$$I_1 = \sum_{T_C\ T'_C\ T''_C\ ,t'_C\ ,t''_C\ 1'_C 1''_C} \mathrm{b}^{00}_{1\,T_C\ 1-T_C}\left\{\sqrt{1/6}\,\mathrm{b}^{1\ T_C}_{00}\ \ \mathrm{b}^{00}_{1\ T'_C\ 1\ T''_C}\,\mathrm{b}^{1\ T_C}_{1T'_C\ 1T''_C}\right.$$
$$\left.+\sqrt{5/6}\,\mathrm{b}^{1\,T_C}_{21'_C\ 11''_C}\,\mathrm{b}^{21'_C}_{1\,T'_C\ 1\ T''_C}\ \mathrm{b}^{11''_C}_{1,t'_C\ 1,t''_C}\right\}\left[u^T_{T'_C}\,C\gamma_\mu\ d_{\,T''_C}\right]\left[u^T_{,t'_C}C\gamma_5 d_{,t''_C}\right]\bar{s}_{-\,T_C}. \tag{14}$$

$$I_2 = \sum_{T_C\ T'_C\ T''_C\ ,t'_C\ ,t''_C\ 1'_C 1''_C} \mathrm{b}^{00}_{1\,T_C\ 1-T_C}\left\{\sqrt{1/6}\,\mathrm{b}^{1\ T_C}_{00}\ \ \mathrm{b}^{00}_{1\ T'_C\ 1\ T''_C}\,\mathrm{b}^{1\ T_C}_{1T'_C\ 1T''_C}\right\}$$
$$\left[u^T_{T'_C}\,C\gamma_5\ d_{\,T''_C}\right]\left[u^T_{,t'_C}C\ \ d_{,t''_C}\right]\bar{s}_{-\,T_C}. \tag{15}$$

Considering flavour configuration of $q^4$, one can see that there are $15_1$ and $15_2$ multiplets which coms from $6\otimes 6$ normal products and this leads to two 45 multiplets for flavour configurations of $q^4\bar{q}$ in which there is octet, decuplet , 27 plet and 35 plet.

The tensor notations of $15_1$ and $15_2$ are $T_{jklm}$ and $S^i_{jk}$ respectively

$$T_{jklm} = \frac{1}{\sqrt{6}}\left(S_{jk}S_{lm}+S_{lk}S_{jm}+S_{jm}S_{kl}+S_{lj}S_{km}+S_{km}S_{jl}+S_{lm}S_{jk}\right),$$
$$S^i_{jk} = \frac{1}{\sqrt{2}}\epsilon^{ilm}\left(S_{jl}S_{km}+S_{kl}S_{jm}\right), \tag{16}$$

Thus by multiplying to tensor notation of $\bar{q}$ ,$T^i$ we leads to tensor notations of ( 8 , 10 , 27 , 35 ) plets.

In fact there is noting in quark model to prevent us for constructing such multiplets in which they have vector diquarks and the discovery of them will be evidence supporting the diquark model.

Briefly the spin - flavour - color and parity of our model for the first version and second one are as follows:

$$\left|\{QQ\}^{\ell=1,\mathbf{3}_c,\bar{\mathbf{6}}_f}\ \bar{q}^{\ j=\frac{1}{2},\bar{\mathbf{3}}_c,\bar{\mathbf{3}}_f}\right\rangle^{J^\Pi=(\frac{1}{2}^+\oplus\frac{3}{2}^+)\ ,\mathbf{1}_c,(\overline{\mathbf{10}}_f\oplus\mathbf{8}_f)} \tag{17}$$

$$\left|\{QQ\}^{\ell=0,\mathbf{3}_c,\bar{\mathbf{6}}_f}\ \bar{q}^{\ j=\frac{1}{2},\bar{\mathbf{3}}_c,\bar{\mathbf{3}}_f}\right\rangle^{J^\Pi=(\frac{1}{2}^-\oplus\frac{3}{2}^-),\mathbf{1}_c,(\mathbf{8}_f\oplus\overline{\mathbf{10}}_f)} \tag{18}$$

We have considered $[\,4\,]^{fs}_{126}$ and $[\,31\,]^{fs}_{210}$ for the flavour - spin configurations of $q^4$ in the first and second version of our model respectively , this leads to $[\,51111]_{700}$ and $[\,42111]_{1134}$ for the flavour - spin configurations of $q^4\bar{q}$ , but if one assume the angular momentum $\ell = 1$ for the four quarks $q^4$ there are several allowed $SU_{fs}(\,6\,)$ representations for $q^4\bar{q}$ which are $[\,51111\,]_{700}$ $[\,42111\,]_{1134}$ $[\,33111\,]_{560}$, $[\,32211\,]_{540}$ based on $[\,4\,]^{fs}$, $[\,31\,]^{fs}$ , $[\,22\,]^{fs}$ and $[\,211\,]^{fs}$ $SU_{fs}(\,6\,)$ representations for $q^4$ respectively.

In order to get an idea of the general features of the spectrum we have calculated the mass spectrum of exotic pentaquark states with the Gürsey - Radicati mass formula [9]:

$$M = M_0 + M_{\rm orb} + M_{\rm sf}\ . \tag{19}$$

in which:

$$M_{\rm sf} = -A\,C_{2SU_{\rm sf}(6)} + B\,C_{2SU_{\rm f}(3)} \pm C\,s(s+1) + D\,Y + E\left[I(I+1)-\frac{1}{4}Y^2\right]. \tag{20}$$

The first two terms represent the quadratic Casimir operators of $SU_{fs}(\,6\,)$ spin - flavour and $SU_f(\,3\,)$ flavour groups and S , Y , and I denote the spin , hypercharge and isospin respectively.

The last two terms in Eq (20) correspond to the Gell- Mann - Okubo mass formula that describes the splitting with in a flavour multiplet.

Neglacting the hyperfine interaction radial dependence the matrix element of these interactions depends on the Casimirs of the $SU_{fs}(6)$ , $SU_f(3)$ and the $SU_s(2)$ [10] groups , also the spatial dependence of the $SU_{fs}(6)$ breaking part has been neglected in Eq (20) [11] .

The cofficients B , C , D and E are determined from the three - quark spectrum [12].

B = 21.2Mev , C = 38.3Mev , D = - 197.3Mev , E = 38.5Mev .

The eigenvalues of the Casimirs for the qqq or $qqqq\bar{q}$ systems are as follows:

Table1

Eigenvalues of the $C_{2SU_{\rm sf}(6)}$ and $C_{2SU_{\rm f}(3)}$ Casimir operators

| spin-flavour | $C_{2SU_{\rm sf}(6)}$ | flavour | $C_{2SU_{\rm f}(3)}$ |
|---|---|---|---|
| $[51111]_{700}$ | 81/4 | $[51]_{35}$ | 12 |
| $[411111]_{56}$ | 45/4 | $[42]_{27}$ | 8 |
| $[42111]_{1134}$ | 65/4 | $[33]_{10}$ | 6 |
| $[321111]_{70}$ | 33/4 | $[411]_{10}$ | 6 |
| $[33111]_{560}$ | 57/4 | $[321]_{8}$ | 3 |
| $[32211]_{540}$ | 49/4 | $[222]_{1}$ | 0 |
| $[222111]_{20}$ | 21/4 | | |
| $[22221]_{70}$ | 33/4 | | |

If we neglect $M_{\rm orb}$ and $A\,C_{2SU_{\rm sf}(6)}$ in Eq (19) and using $M_0$ in order to normalize the energy scale to the observed mass of the $\Theta^+$ ( 1540 ),

We find that a flavour anti-decuplet $[\,33\,]^{f}_{\overline{10}}$ state with spin S=1/2 and isospin I = 0 which has $[\,4\,]^{fs}_{126}$ $q^4$ flavour - spin configuration based on q $[\,22\,]^{f}$ flavour and $[\,22\,]^{s}$ spin configuration and lead to $[51111]^{fs}$ for $q^4\bar{q}$ which we have introduced in the first version of our model is the lowest pentaquark state.

The state is in agreement with the available experimental data which indicate that $\Theta^+$ ( 1540 ) is an isosingle, this means that the contributions of $M_{\rm orb}$ and $A\,C_{2SU_{\rm sf}(6)}$ $SU_{fs}(6\,)$ terms in Eq (19) must be camparable and this is a good test for the values of exciting energy of the orbital degrees of freedom for $q^4$ and the underlying dynamical structure.

However , one can see by neglecting this two terms the mass spectrum of al $[\,51111]^{fs}_{700}$ $[\,33111]^{fs}_{560}$ $[\,32211]^{fs}_{540}$ $q^4\bar{q}$ multiplets are degenerate, this is of because that the only term which split the mass spectrum between these configurations and depends on the flavour- spin structure of them is $A\,C_{2SU_{\rm sf}(6)}$ in general the coficients C, D ,E depends on the internal structure of pentaquark and we are not allowed to use of the corresponding three quark values.

The hyperfine interactions between two quarks in diquarks and the confinement between quarks are the underlying interactions which have ignored in this general study of the structure of the spectrum.

We have mentioned that one find ( 8 , 10 , 27 , 35 ) multiplets by decomposition of 45 multiplets for flavour configurations of $q^4\bar{q}$.

The $SU_f(\,3\,)$ configurations of these multiplets are $[\,321\,]_8$, $[\,411\,]_{10}$ $[\,42\,]_{27}$ and $[\,51\,]_{35}$ and the $SU_{fs}(\,6\,)$ configurations of them are :

$[\,33111]_{560}$, $[\,321111]_{70}$, $[\,41111]_{56}$ and $[\,51111]_{700}$

We do not give the full list of these pentaquark masses and one can read them directly from the results presented in Table .1 for $C_{2SU_f(3)}$ and $C_{2SU_{sf}(6)}$ and by finding their quantum numbers [6] according to tensor notations of them and using of Eq (20) for their masses.

However, observation of other pentaquark states in higher multiplets will help us to understand the structure of pentaquark baryons specially diquark- diquark - antiquark models.